\documentclass{ieeeaccess}
\usepackage{cite}
\usepackage{amsmath,amssymb,amsfonts}
\usepackage{graphicx}
\usepackage{textcomp}

\usepackage{mathtools}

\usepackage{algorithm}
\usepackage{algpseudocode}
\algnewcommand{\Initialize}[1]{%
  \Statex \textbf{Initialize:}
  \Statex \hspace*{\algorithmicindent}\parbox[t]{.8\linewidth}{\raggedright #1}
}
\algnewcommand\algorithmicbegin{\textbf{Begin}}
\algnewcommand\Begin{\item[\algorithmicbegin]}
\algnewcommand\algorithmicendd{\textbf{End}}
\algnewcommand\Endd{\item[\algorithmicendd]}

\usepackage{caption,setspace}


\def\BibTeX{{\rm B\kern-.05em{\sc i\kern-.025em b}\kern-.08em
    T\kern-.1667em\lower.7ex\hbox{E}\kern-.125emX}}

\begin{document}
\history{PREPRINT. WORK IN PROGRESS.}
\doi{10.1109/DOI}

\title{Wasserstein GAN and Waveform Loss-based Acoustic Model Training for Multi-speaker Text-to-Speech Synthesis Systems Using a WaveNet Vocoder}
\author{\uppercase{Yi Zhao}\authorrefmark{1}, \IEEEmembership{Student Member, IEEE},
\uppercase{Shinji Takaki}\authorrefmark{2}, \IEEEmembership{Member, IEEE},\\
\uppercase{Hieu-Thi Luong}\authorrefmark{2},\IEEEmembership{Student Member, IEEE},
\uppercase{Junichi Yamagishi}\authorrefmark{2,3},\IEEEmembership{Senior Member, IEEE},
\uppercase{Daisuke Saito}\authorrefmark{1},\IEEEmembership{Member, IEEE},
\uppercase{Nobuaki Minematsu}\authorrefmark{1},\IEEEmembership{Member, IEEE}
}

\address[1]{Department of Electrical Engineering and Information Systems, Graduate School of Engineering, The University of Tokyo, 7-3-1 Hongo, Bunkyo-ku, Tokyo 113-8656, Japan (e-mail: zhaoyi@gavo.t.u-tokyo.ac.jp, dsk\_saito@gavo.t.u-tokyo.ac.jp, mine@gavo.t.u-tokyo.ac.jp)}
\address[2]{Digital Content and Media Sciences Research Division, National Institute of Informatics, 2-1-2 Hitotsubashi, Chiyoda-ku, Tokyo 101-8430, Japan (e-mail: takaki@nii.ac.jp, luonghieuthi@nii.ac.jp, jyamagis@nii.ac.jp)}
\address[3]{The Centre for Speech Technology Research, University of Edinburgh, 10 Crichton Street, EDINBURGH, EH89AB, United Kingdom}




\tfootnote{This work was partially supported by MEXT KAKENHI Grant Numbers 16H06302, 17H04687, 18H04120, and 18H04112. The authors thank Mr.\ Lauri Juvela from Aalto University, Finland for his suggestions on WGAN-GP.}

\markboth
{Yi Zhao \headeretal: Wasserstein GAN and Waveform Loss-based Acoustic Model Training for Multi-speaker TTS Systems}
{Yi Zhao \headeretal: Wasserstein GAN and Waveform Loss-based Acoustic Model Training for Multi-speaker TTS Systems}

\corresp{Corresponding author: Yi Zhao(e-mail: zhaoyi@gavo.t.u-tokyo.ac.jp).}

\begin{abstract}
Recent neural networks such as WaveNet and sampleRNN that learn directly from speech waveform samples have achieved very high-quality synthetic speech in terms of both naturalness and speaker similarity even in multi-speaker text-to-speech synthesis systems. Such neural networks are being used as an alternative to vocoders and hence they are often called neural vocoders. The neural vocoder uses acoustic features as local condition parameters, and these parameters need to be accurately predicted by another acoustic model. However, it is not yet clear how to train this acoustic model, which is problematic because the final quality of synthetic speech is significantly affected by the performance of the acoustic model. Significant degradation happens, especially when predicted acoustic features have mismatched characteristics compared to natural ones. In order to reduce the mismatched characteristics between natural and generated acoustic features, we propose frameworks that incorporate either a conditional generative adversarial network (GAN) or its variant, Wasserstein GAN with gradient penalty (WGAN-GP), into multi-speaker speech synthesis that uses the WaveNet vocoder. We also extend the GAN frameworks and use the discretized mixture logistic loss of a well-trained WaveNet in addition to mean squared error and adversarial losses as parts of objective functions. Experimental results show that acoustic models trained using the WGAN-GP framework using back-propagated discretized-mixture-of-logistics (DML) loss achieves the highest subjective evaluation scores in terms of both quality and speaker similarity.
\end{abstract}


\titlepgskip=-15pt

\maketitle
\section{Introduction}
\label{sec:intro}
In recent years, text-to-speech (TTS) synthesis has gained popularity as an artificial intelligence technique and is widely used in many applications with speech interfaces. There are currently two major categories in the machine learning-based speech synthesis field: a) an end-to-end approach that learns the relationship between text and speech directly and b) the conventional pipeline processing approach that divides text-to-speech conversion into sub tasks such as linguistic feature extraction and acoustic feature extraction. In the latter approach, an acoustic model is trained to learn the relationship between separately extracted linguistic and acoustic features~\cite{black2007statistical}. Previously investigated acoustic models include the hidden Markov model (HMM)~\cite{ze2013statistical}, the deep neural network (DNN)~\cite{zen2014deep}, and the recurrent neural network (RNN)~\cite{fan2014tts}\cite{zen2015unidirectional}. These are normally trained with the minimum mean squared error (MSE) criterion, and hence, the generated acoustic parameters tend to be over-smoothed regardless of the architectures. Finally, speech waveforms have been reconstructed using a deterministic vocoder based on the acoustic parameters~\cite{erro2014harmonics}\cite{agiomyrgiannakis2015vocaine}\cite{espic2017direct}. However, the generated signals have artifacts and typically sound buzzy. Due to these two major issues, the resultant quality of generated speech sounds obviously worse compared with natural speech.

Very recently, we see emerging solutions for the two issues. To alleviate the over-smoothing problem, Saito et al.\ have incorporated adversarial training into acoustic modeling~\cite{saito2017training}\cite{saito2018statistical}. The generative adversarial network (GAN) contains a generator as well as a discriminator~\cite{goodfellow2014generative}, where the generator aims at deceiving the discriminator and the discriminator is trained to distinguish the natural and generated feature samples. In the framework proposed by~\cite{saito2018statistical}, the generator acts as an acoustic model and is optimized by not only the conventional MSE but also an adversarial loss computed using the discriminator. Experimental results show that GAN can effectively alleviate the over-smoothing effect of the generated speech parameters.

To avoid the artifacts and deterioration caused by deterministic vocoders, WaveNet, which directly models the raw waveform of the audio signal in a non-linear auto-regressive way, has been proposed and dramatically improves the quality of synthetic speech~\cite{van2016wavenet}\cite{oord2017parallel}. The original WaveNet model~\cite{van2016wavenet} used linguistic features as well as the fundamental frequency (F0) as local conditions. Later, the WaveNet model was used as an alternative to the deterministic vocoders in many studies~\cite{tamamori2017speaker}\cite{shen2017natural} by conditioning it on acoustic features such as cepstrum, F0, or spectrograms only~\cite{tamamori2017speaker}, and results have shown that the sound quality of the WaveNet vocoder outperformed deterministic vocoders and phase recovery algorithms~\cite{wang2018comparison}.

However, it is also reported that the samples generated from WaveNet occasionally become unstable and generate collapsed speech, especially when less accurately predicted acoustic features are used as the local condition parameters~\cite{wu2018collapsed}. This would be more critical for the case of multi-speaker acoustic modeling where the same network is used for modeling multiple speakers at the same time, as the prediction accuracy of the multi-speaker model would be worse than well-trained speaker-dependent models. 

In this paper, we propose frameworks that incorporate either the conditional GAN~\cite{mirza2014conditional} or its variant, Wasserstein GAN with gradient penalty (WGAN-GP)~\cite{gulrajani2017improved}, into RNN-based speech synthesis systems using the WaveNet vocoder for the purpose of reducing the mismatched characteristics between natural and generated acoustic features and for making the outputs of the WaveNet vocoder better and more stable. We evaluate the proposed frameworks using a multi-speaker modeling task. The generator of GAN is conditioned on both linguistic features and speaker code, and the discriminator aiming at distinguishing the real and predicted mel-spectrograms is also conditioned on speaker information. The WaveNet vocoder is conditioned on both mel-spectrogram and speaker codes, as well.

In addition, we extend the GAN frameworks and define a new objective function using the weighted sum of three kinds of losses: conventional MSE loss, adversarial loss, and discretized mixture logistic loss~\cite{salimans2017pixelcnn++} obtained through the well-trained WaveNet vocoder. Since the third loss will let neural networks consider losses not only in the acoustic feature domain (such as mel-spectrogram) but also in the final waveform, we hypothesize that it will improve the quality of synthetic speech. In our experiment, simple recurrent units (SRUs)~\cite{lei2017training} are utilized as basic components since they can be trained faster than the LSTM-based RNN architecture while maintaining a performance as good as or even better than LSTM-RNN.

In Section 2 of this paper, we briefly review previously proposed DNN-based multi-speaker speech synthesis, as we evaluate our proposed method in a popular multi-speaker modeling task. In Section~\ref{sec:gan_wav}, we present the proposed framework for multi-speaker speech synthesis. Section~\ref{sec:relate} describes the basic elements of the structure of the proposed model including SRU, GAN, and WaveNet, and the details of the training algorithms are given in Section~\ref{sec:alg}. Section~\ref{sec:exp} describes experimental conditions and Section~\ref{sec:res} discusses the results. We conclude in Section~\ref{sec:con} with a brief summary and mention of future work.

\section{DNN-based Multi-speaker Speech Synthesis} 
Although deep learning-based methods have significantly advanced the performance of statistical parametric speech synthesis (SPSS), it still suffers from the necessity of a large amount of speech recordings of one speaker to train a high-quality acoustic model. Ideally, a speech synthesis system should be able to generate an arbitrary speaker's voice with a minimum of training data. Multi-speaker speech synthesis is one of the most effective approaches to train such a high-quality acoustic model with a limited amount of speech data of each speaker. Using multiple speakers' data at the same time, we can improve the quality of synthesized speech and can also change the speaker characteristics of synthetic speech flexibly.

Using DNN-based acoustic models as a basis, Fan et al.~\cite{fan2015multi} proposed multi-speaker speech synthesis using shared speaker-independent layers as well as a speaker-dependent output layer. They showed that the speaker-dependent output layer can be estimated from a target speaker's data only and that the shared hidden layers can improve the quality of synthesized speech of individual speakers. Wu et al.~\cite{wu2015study} suggested using i-vectors for modeling multiple speakers and controlling the speaker identity of synthetic speech. Hojo et al.~\cite{hojo2016investigation} proposed using speaker codes based on a one-hot vector for modeling multiple speakers and extending the code and associated weights at an input layer for adapting it to unseen speakers. Luong et al.~\cite{luong2017adapting} proposed estimating code vectors for new speakers via back-propagation and experimented with manually manipulating input code vectors to alter the gender and/or age characteristics of the synthesized speech. Similar work has been extended to LSTM-based acoustic models. Zhao et al.~\cite{zhao2016speaker} examined various speaker identity representations for multi-speaker synthesis and showed that multi-speaker systems trained with less of the target speaker's data can even outperform single speaker speech synthesis, which uses a larger amount of the target speaker's data. Li et al.~\cite{li2016multi} investigated multi-speaker modeling with speech data in different languages.

Multi-speaker speech synthesis has also been investigated in the recent WaveNet-based approaches and in end-to-end approaches. Hayashi~\cite{hayashi2017investigation} attempted WaveNet vocoder-based multi-speaker synthesis using four speakers from the CMU arctic corpus~\cite{kominek2004cmu}. VoiceLoop~\cite{taigman2018voiceloop} involves the data of 109 speakers for acoustic model training, and Deep Voice 3~\cite{ping2018deep} trained a multi-speaker model using over 2,000 speakers. Wang et al.~\cite{wang2018style} proposed a bank of style embedding vectors and used it for modeling multiple TED speakers. As we can see, very active research on multi-speaker modeling has been carried out.

\section{Multi-speaker Speech Synthesis Incorporating GAN and WaveNet Vocoder}
\label{sec:gan_wav}
In this section, we introduce the proposed speech synthesis framework for multi-speaker modeling.

\begin{figure}[tb]
\centering
\includegraphics[clip, trim=2cm 7cm 2.5cm 2.5cm, width=0.5\textwidth]{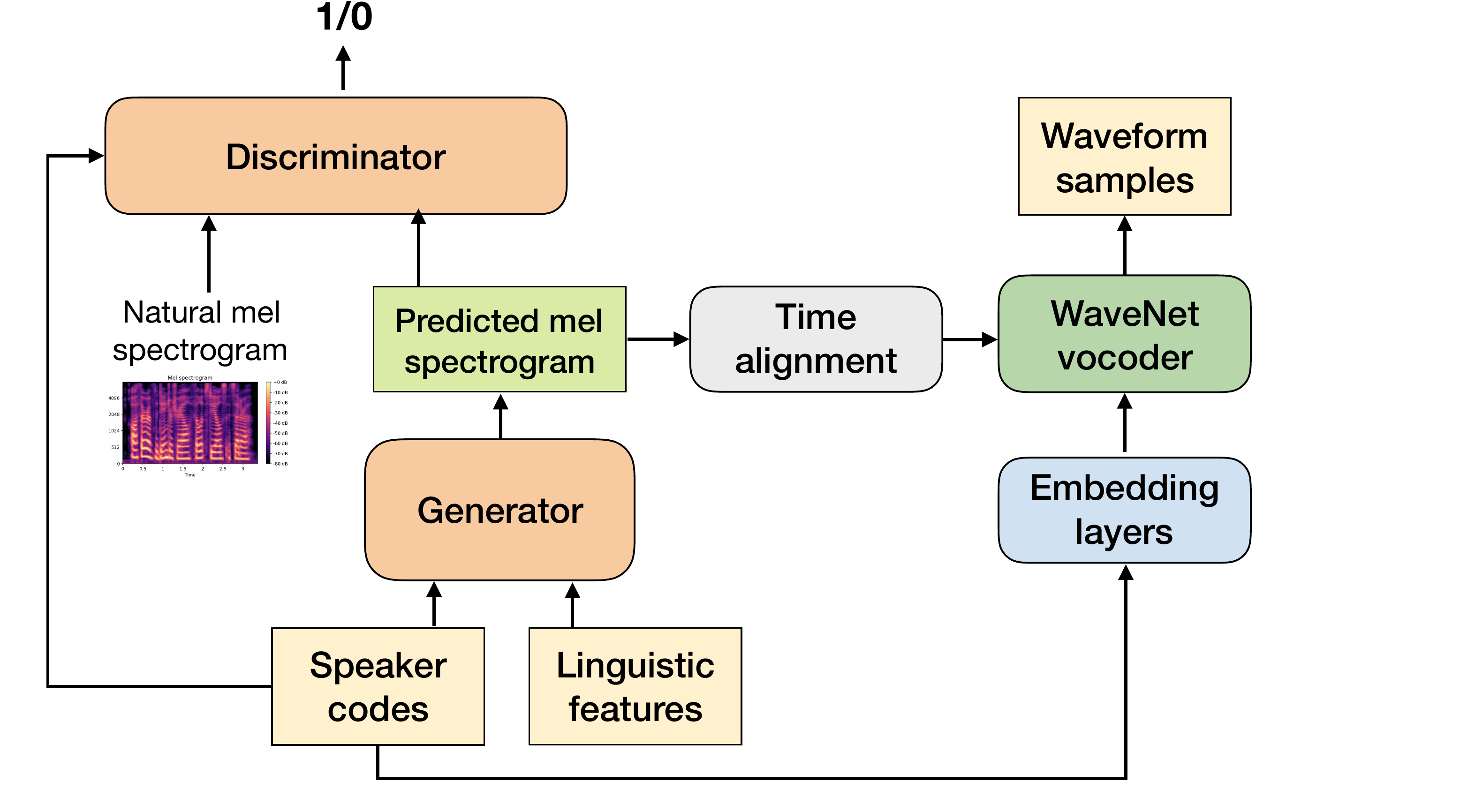} \\
\caption{Proposed GAN-trained multi-speaker speech synthesis framework using a WaveNet vocoder.}
\label{img:frame}
\end{figure} 

In the conventional SPSS structure, acoustic models and vocoders usually work independently: the acoustic models are trained without any consideration of the speech vocoding process, and vice versa. It was the same in the first versions of end-to-end structures such as Deep Voice~\cite{arik2017deep}, where vocoders were usually designed or trained on natural acoustic parameters without considering the divergence between predicted and natural acoustic parameters. This may lead to obvious and unpredictable distortion of the synthesized speech. To alleviate this problem, Tacotron 2~\cite{shen2017natural} utilized predicted mel-spectrograms to train the WaveNet vocoder instead of natural mel-spectrograms. Experimental results showed that such a strategy may outperform those that use natural parameters and may achieve a higher evaluation. 

In the present work, we try to minimize the acoustic mismatch of predicted and natural parameters by conducting acoustic model training based on GAN, which also considers vocoder loss. The proposed multi-speaker speech synthesis framework is shown in Fig.~\ref{img:frame}. In this framework, a generator part of GAN is adopted to predict acoustic features from linguistic features, and both the generator and discriminator are conditioned on speaker codes and trained with multiple speakers' data. Similar to Tacotron 2, the mel-spectrogram, a low-dimensional representation of the linear-frequency spectrogram, which contains both spectral envelop and harmonics information, is selected as the output of the generator and used to bridge the acoustic model and the WaveNet vocoder. Mel-scale acoustic features have overwhelming advantages in terms of emphasizing the details of audio, especially for lower frequencies, since they are more critical to phonetic information and hence to speech intelligibility in general. 

The input of the discriminator is either natural or generated acoustic feature samples. The discriminator is trained to distinguish natural samples from generated ones. Speaker codes are also attached to both the input and hidden layers of the discriminator in order to make a better distinction between different speakers. The discriminator is used to compute the adversarial (ADV) loss, which is expected to alleviate the over-smoothing problem. 

In addition to the adversarial (ADV) loss from the discriminator, the average discretized-mixture-of-logistics (DML) loss of a well-trained WaveNet model is also back-propagated to the generator of GAN. This loss corresponds to distortion between natural and generated waveform samples. We hypothesize that this increases the consistency of acoustic features predicted by the acoustic model and utilized in the vocoder since the acoustic model is updated on the basis of gradients directly computed by the pre-trained WaveNet vocoder. 

In brief, it is expected that the weighted sum of the conventional MSE loss, the adversarial loss of the discriminator, and the DML from the WaveNet vocoder will improve the accuracy of the predicted acoustic parameters and thus enhance synthesized speech quality. What sets this work apart from other related works is that WaveNet is involved in the process of acoustic modeling training. After extracting acoustic features from a training corpus, the WaveNet vocoder is first trained by utilizing natural mel-spectrograms, and then the trained WaveNet model is fixed and referenced for acoustic model optimization.

\section{Components of Proposed Model Structure}
\label{sec:relate}
In this section, we describe the three major components of the proposed framework, namely, the SRU architecture and the GAN and WaveNet models.

\subsection{SRU}\label{subsec:SRU}

\begin{figure}[tb]
\begin{center}
\includegraphics[clip, trim=2.5cm 6cm 2.5cm 6.5cm, width=0.5\textwidth]{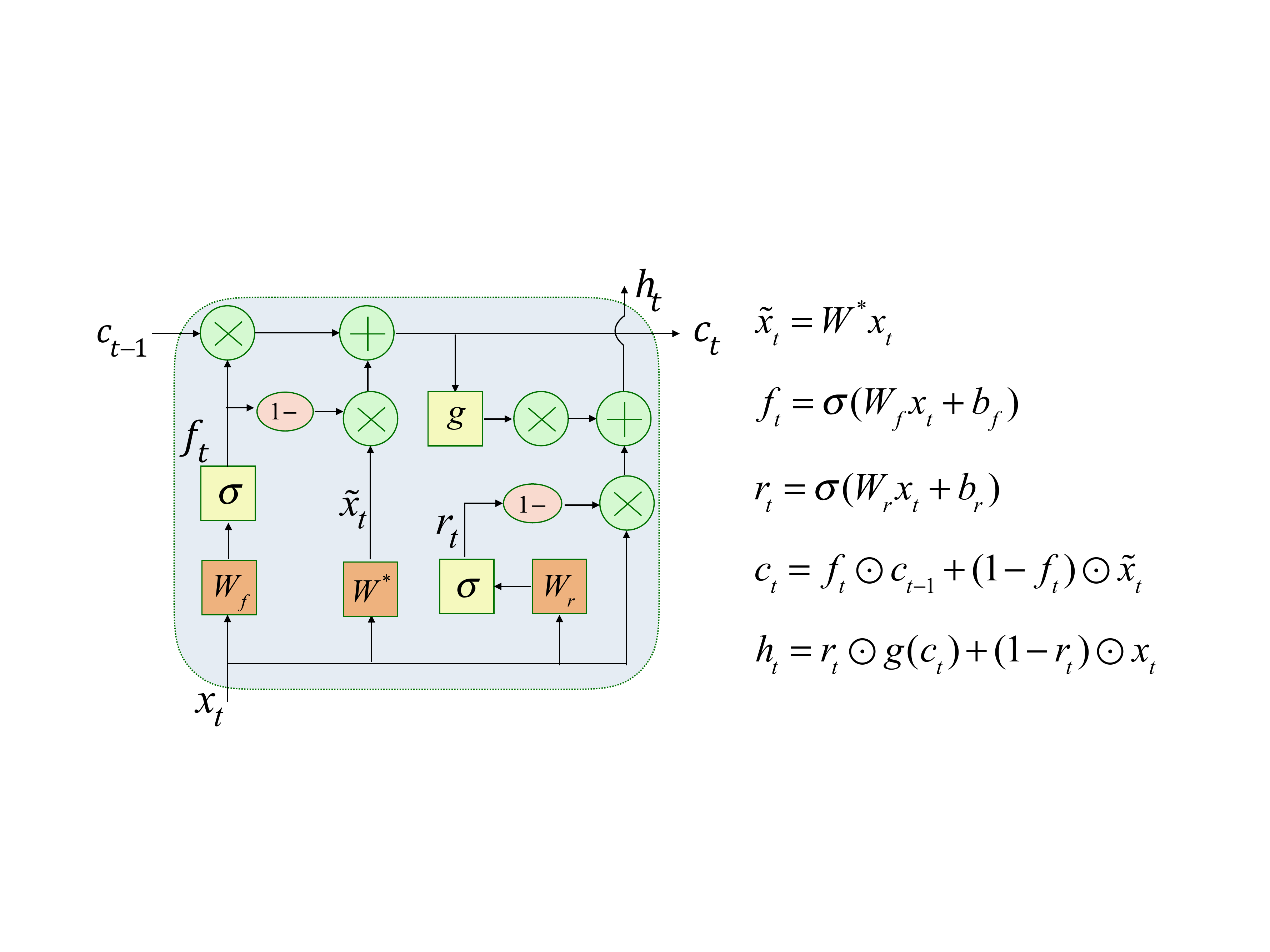}
\end{center}
\caption{Details of the SRU cell. $\sigma(\cdot)$ and $g(\cdot)$ represent sigmoid and ReLU activation functions, respectively.}
\label{img:sru}
\end{figure}

For the sake of modeling accuracy as well as time efficiency, we choose SRU~\cite{lei2017training} as the basic architecture of the acoustic modeling. The SRU architecture was originally designed to speed up the training process of RNN. By utilizing both skip and highway connections, SRU is capable of outperforming RNN, especially on very deep networks. Compared with other recurrent architectures (e.g., LSTM and GRU), the basic form of SRU includes only a single forget gate $f_t$ to alleviate vanishing and exploding gradient problems instead of using many different gates to control the information flow. In SRU, the forget gate is used to modulate the internal state $c_t$, which is then used to compute the output state $h_t$. Unlike existing RNN architectures that use the previous output state in the recurrence computation, SRU completely drops the connection between the gating computations and the previous states, and this makes SRU computationally efficient and allows us to use parallelization. The complete architecture of SRU is shown in Fig.~\ref{img:sru}. The reset gate $r_t$ is computed similar to the forget gate $f_t$ and is used to compute the output state $h_t$, which performs as a combination of the internal state $g(c_t)$ and the input $x_t$. $g(\cdot)$ represents a ReLU activation function and $\sigma(\cdot)$ is a sigmoid function.

\subsection{Generative Adversarial Network}
\label{subsec:gan}
GANs have achieved great success in modeling the distributions of complex data and the predictions of realistic data in many applications. They have also proven beneficial for speaker-dependent speech synthesis~\cite{saito2018statistical}.

\begin{figure}[tb]
\includegraphics[clip, trim=8cm 12cm 2.5cm 3cm, width=0.5\textwidth]{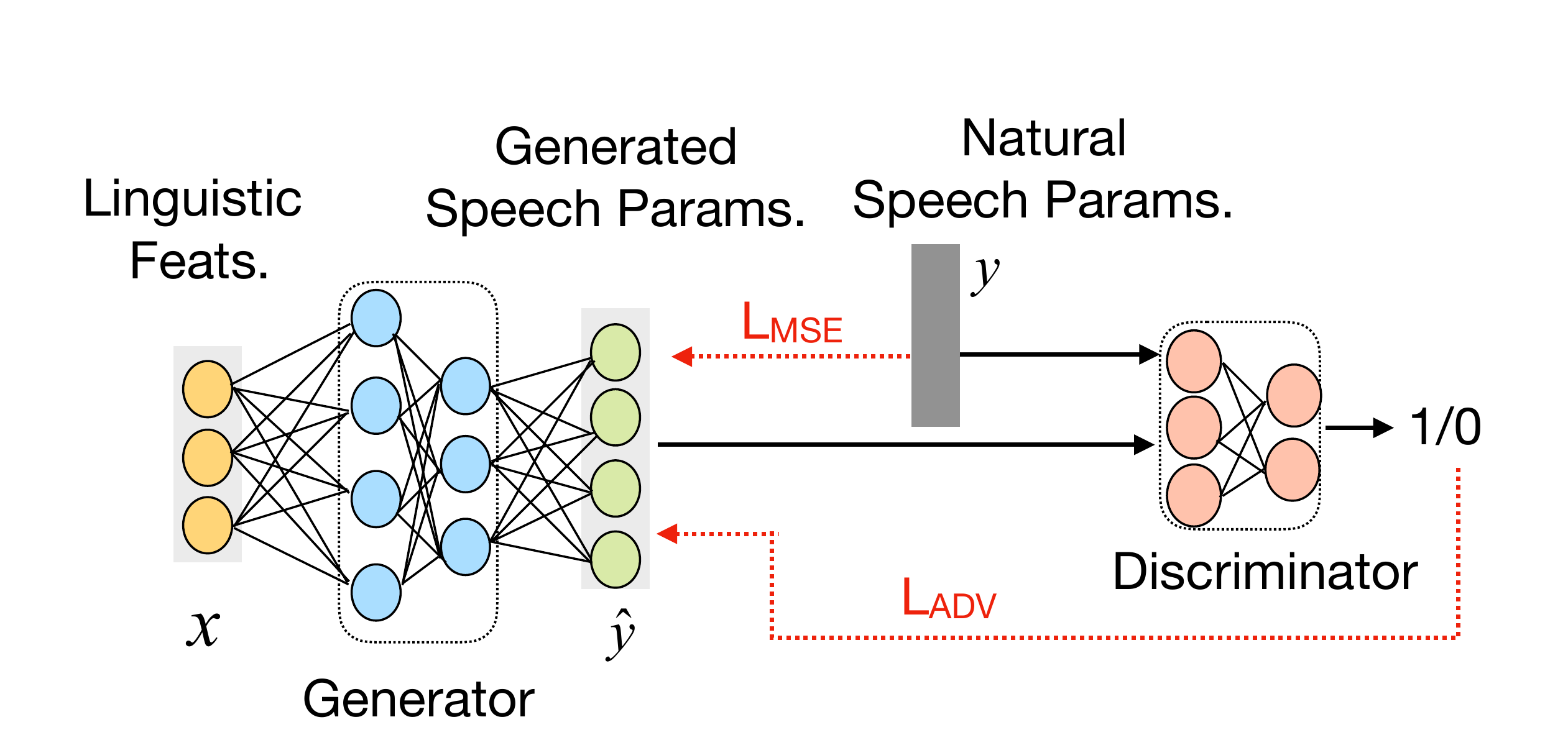} \\
\caption{GAN-based training of TTS acoustic model. $L_{ADV}$ indicates adversarial loss and $L_{MSE}$ indicates L2 loss.}
\label{img:gantts}
\end{figure}

Fig.~\ref{img:gantts} shows the GAN-based training of acoustic models for TTS systems. The GAN training involves a pair of networks: a generator $G$ aims to produce vivid feature samples that deceive a discriminator $D$, and the discriminator aims to estimate the probability that a sample $y$ came from the real data set distribution $\mathbb{P}_r$ rather than a generator distribution $\mathbb{P}_g$. For speech synthesis from text, the generator is conditioned on linguistic vectors $x\sim{\mathbb{P}_x}$. The generator and discriminator are trained like a two-player min-max game objective function, as
\begin{equation}
\min_{G}\max_{D}\underset{{y\sim{\mathbb{P}_r}}}{\mathbb{E}}[\log D(y)] + \underset{{x\sim{\mathbb{P}_x}}}{\mathbb{E}}[\log(1-D(G(x)))]
\end{equation}

This objective function is not easy to optimize. To improve the stability of model training, Wasserstein GAN (WGAN), which minimizes a different distribution divergence called $Earth-Mover$ (EM) or Wasserstein-1 distance, has been proposed and achieved a better performance than original GAN in terms of convergence, especially in image processing~\cite{arjovsky2017wasserstein}. The optimization criteria for WGAN is equal to
\begin{equation}\label{eq:wgan}
\min_{G}\max_{D}\underset{{y\sim{\mathbb{P}_r}}}{\mathbb{E}}[D(y)] - \underset{{x\sim{\mathbb{P}_x}}}{\mathbb{E}}[D(G(x))]
\end{equation}

During the training of WGAN, the updated model parameters of discriminator are clipped into a compact space $[-c, c]$ to enforce a Lipschitz constraint on $D$. However, the weight clipping may lead to either vanishing or exploding gradients if the clipping threshold $c$ is not carefully tuned, and the resulting discriminator may have a pathological value surface even when optimization performs smoothly~\cite{gulrajani2017improved}. To address this problem, Gulrajani et al.~\cite{gulrajani2017improved} proposed penalizing the norm of the gradient deduced from a discriminator with respect to its input. The new objective for WGAN with gradient penalty (WGAN-GP) is shown as follows:
\begin{eqnarray}\label{eq:wgan_gp}
\min_{G}\max_{D}&\underset{{y\sim{\mathbb{P}_r}}}{\mathbb{E}}[D(y)]-\underset{{x\sim{\mathbb{P}_x}}}{\mathbb{E}}[D(G(x))]& \nonumber\\
				 &\mbox{}+
\lambda\underset{{{\tilde{y}}\sim{\mathbb{P}_{\tilde{y}}}}} {\mathbb{E}}[(\left \| \nabla_{\tilde{y}}{D({\tilde{y}})} \right \|_2-1)^2]
\end{eqnarray}
where $\tilde{y}$ represents samples that are linearly interpolated by the real data $y$ and the fake data generated from the generator $G(x)$:
\begin{eqnarray}\label{eq:wgan_gp_ty}
\tilde{y} = \epsilon y+(1-\epsilon)G(x) 
\end{eqnarray}
where $\epsilon$ is a random number that obeys distribution $U[0, 1]$.

The loss function of the generator is also expanded on the basis of the least square errors of $y$ as:
\begin{equation}\label{eq:gan_tts}
L_G(y,\hat{y})=L_{MSE}(y, \hat{y}) + \gamma_D L_{ADV}(\hat{y})
\end{equation}
where $L_{ADV}(\hat{y})$ is the adversarial loss and $\gamma_D$ controls the weight of the adversarial loss. When $\gamma_D = 0$, the loss function is equivalent to the conventional MSE criteria. In original GAN, $L_{ADV}(\hat{y})$ equals $\mathbb{E}[\log(1-D(G(x)))]$. In WGAN-GP, $L_{ADV}(\hat{y})$ can be regarded as $-\mathbb{E}[D(G(x))]$. 

\subsection{WaveNet}
\label{sec:wavnet}
\begin{figure}[tb]
\centering
\includegraphics[clip, trim=2cm 15cm 0cm 2.5cm, width=0.5\textwidth]{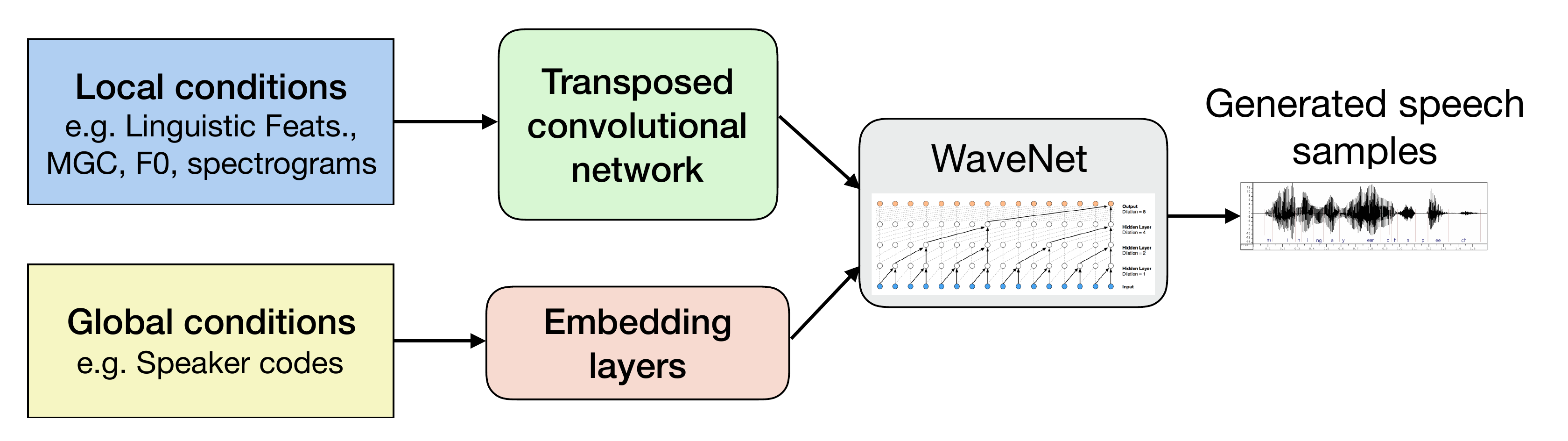} \\
\caption{Local condition and global condition used in a WaveNet model.}\label{img:condition}
\end{figure}

WaveNet is a deep auto-regressive and generative model that models a joint distribution of sequential data as a product of conditional distributions, as
\begin{equation}
\centering
p(s)=\prod _{t}p(s_t|s_{<t}, \theta)
\end{equation}
where $s_t$ is a variable of $s$ at time $t$ and $\theta$ denotes model parameters. The conditional distributions are usually modelled with a neural network that receives all past variables $s_{<t}$ as input and outputs a distribution over possible $s_t$. The neural network consists of stacked dilated causal convolution layers~\cite{van2016wavenet}, and each causal convolutional layer can process its input in parallel, making these architectures very fast to train compared to RNNs. It typically uses gated activation functions~\cite{van2016conditional} along with two conditions, global and local, which is another important concept in WaveNet. 

The difference between the two conditions is shown in Fig.~\ref{img:condition}. The global condition focuses on conditional vectors irrelevant to time, e.g.\, a speaker embedding in a TTS model, while the local condition deals with time-series input conditions, such as linguistic and acoustic features. The basic activation function with global conditioning is 
\begin{equation}
\centering
h_i=\sigma (W_{g,i}\ast s_i + V_{g,i}^Tc)\odot\tanh(W_{f,i}\ast s_i + V_{f,i}^Tc) 
\end{equation}
where $\ast$ denotes a convolution operator and $\odot$ denotes an element-wise multiplication operator. $\sigma$ is a logistic sigmoid function. $c$ represents a global condition. $i$ is the layer index. $f$ and $g$ denote filter and gate, respectively. $W$ and $V$ are learnable weights. For a case where $c$ denotes the local condition (such as mel-spectrogram), the matrix products $V_{g,i}^Tc$ and $V_{f,i}^Tc$ are replaced by convolutions $V_{g,i}^T\ast c$ and $V_{f,i}^T\ast c$, respectively.

Oord et al.~\cite{van2016wavenet} take both linguistic and acoustic features such as F0 as the local conditions. In other studies~\cite{tamamori2017speaker,ping2018deep,shen2017natural}, only acoustic features are used as the local conditions, and the WaveNet model tends to perform as a neural vocoder. In the proposed framework, WaveNet is used as a multi-speaker neural vocoder. It is locally conditioned on mel-spectrograms and globally conditioned on speaker embeddings.

\subsection{DML loss}
\label{subsec:wavnet}
In~\cite{van2016wavenet}, speech waveform samples were quantized and the cross entropy loss was used for modeling categorical distribution, but if we use additional quantization bits (to reduce the quantization noise), the cost of computations may be exponentially increased. Using discretized mixture of logistics (DML) distribution loss~\cite{salimans2017pixelcnn++} could save memory and improve training efficiency because it just needs to predict parameters for each mixture component instead of all bits. For example, modeling 16-bit quantized bits always requires the training of a 65,536-way categorical distribution, while only ten mixtures of logistic distributions are sufficient to model 16-bit audio samples empirically.

DML distribution assumes that each sample point $s$ is composed of a mixture of continuous uni-variate distributions $\upsilon$, and each component $\upsilon_i$ obeys logistic distribution, as 
\begin{equation}\label{eq:v_i}
\upsilon =\sum_{i=1}^K{\pi_i v_i}, \qquad \text{where } \upsilon _i\sim{\text{logistic}(\mu_i,\phi_i)}
\end{equation}
where $\pi_i$ is the mixture weight of component $i$ that satisfies $\sum_{i=1}^{K}\pi_i=1$. $\mu$ is the mean and $\phi$ is a scale parameter proportional to the standard deviation. The probability on the observed discretized audio sample $s$ excepting the edge cases (e.g., 0 and 65,535 for 16-bit sampling) would be
\begin{equation}\label{eq:pos_s}
P(s|\pi,\mu,\phi)=\sum_{i=1}^K{\pi_i[\sigma(\frac{s+1-\mu_i}{\phi_i\zeta}) - \sigma(\frac{s-1-\mu_i}{\phi_i\zeta})]}
\end{equation}
$\sigma(\cdot)$ is the logistic sigmoid function. $\zeta$ denotes the number of sampling classes and $\zeta = 256$ for 8-bit and $65536$ for 16-bit sampling. For the edge case of 0, replace $s-1$ with $-\infty $, and for 255 or 65535, replace $s+1$ with $+\infty$. Finally, the WaveNet model aims at maximizing the average log likelihood of $P$:
\begin{equation}\label{eq:DML}
L_{DML} = \max_{W}{\mathbb{E}[\log(P(s|\hat{\pi},\hat{\mu},\hat{\phi}))]}
\end{equation}
where $\hat{\pi},\hat{\mu},\hat{\phi}$ are predicted mixture component parameters.

\section{Training Algorithm}
\label{sec:alg}
\subsection{Training algorithm for the proposed acoustic model}
\begin{figure}[tb]
\begin{center}
\includegraphics[clip, trim=9.5cm 8.5cm 3cm 5cm, width=0.45\textwidth]{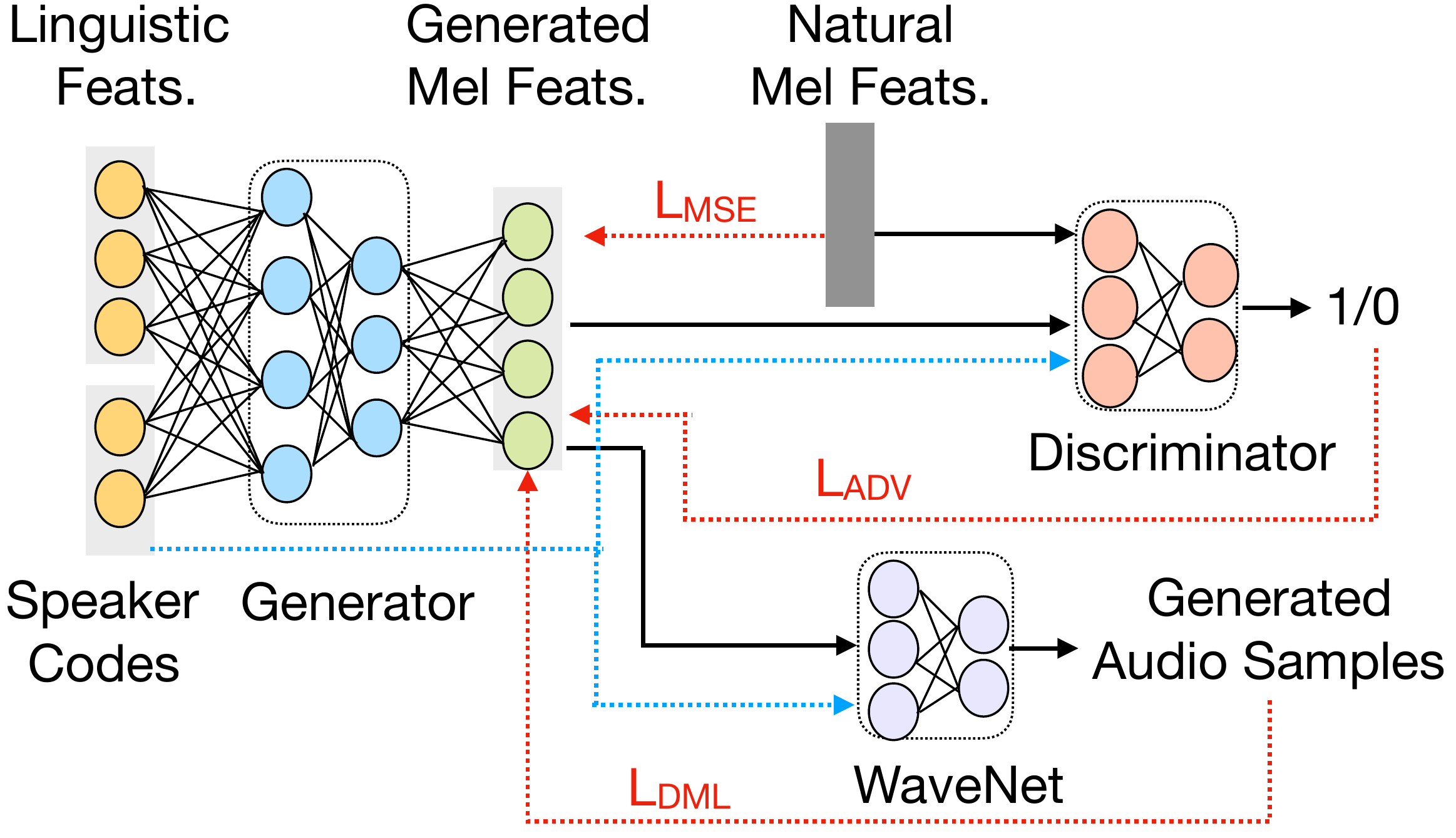}
\end{center}
\caption{Loss functions and gradients for updating acoustic models in the proposed method. Note that neither the model parameters of WaveNet nor the discriminator are updated in this step.}
\label{img:loss_back}
\end{figure}   

The overall loss function for training the proposed acoustic model that predicts mel-spectrogram can be written as
\begin{equation}\label{eq:gan_wav_tts}
L_{G}(y, \hat{y}) = L_{MSE} (y,\hat{y}) + \gamma_D L_{ADV} (\hat{y}) + \gamma_W L_{DML} (y,\hat{y})
\end{equation}
In addition to the general MSE loss $L_{MSE}$ and adversarial loss $L_{ADV}$, the DML loss $L_{DML}$ generated by a well-trained WaveNet model is utilized for updating the model parameters of the generator. Utilizing the DML loss with the generator would integrate the divergence of synthesized speech samples into the acoustic parametric training process. Therefore, the proposed loss function minimizes not only the parametric error of the mel-spectrogram but also the fidelity disparity between predicted and natural audios. $\gamma_W$ is a hyper-parameter that denotes the weight of $L_{DML}$. When $\gamma_W$ = 0, the loss function is equivalent to the conventional GAN training. Model parameters of the generator $\theta_G$ are updated by using the stochastic gradient calculated from $L_G(y, \hat{y})$. Fig.~\ref{img:loss_back} shows the procedure for computing the proposed loss function.

The details of the acoustic model training algorithm are given in Algorithm~\ref{al:train}. In the first step, the generator is trained with the MSE criterion for a few epochs. Then, the generator and discriminator are optimized in an iterative way, where one module is being updated while the model parameters of another are fixed. In the final step, the loss of WaveNet $L_{DML} (y,\hat{y})$ is enrolled in the training criterion of the generator. Before this step, the WaveNet vocoder needs to be trained in advance and the optimum model parameters $\theta_{W}$ should be saved. Note that the DML loss does not join the optimization process of the discriminator, and the parameters of the WaveNet model are always kept fixed. In other words, although $\theta_D$ and $\theta_W$ are included in calculating $L_G(y,\hat{y})$, $\theta_D$ is not updated by the back-propagation of $L_G$ in the final step, and neither is $\theta_W$. The WaveNet model is used as a measurement that reflects the divergence between speech samples. In the WGAN-GP-based case, $\theta_D$ is first optimized according to Eq. (\ref{eq:wgan_gp}) and then $\theta_G$ is optimized according to Eq. (\ref{eq:gan_wav_tts}).

\begin{algorithm}
\caption{Training algorithm for acoustic modeling.}
\label{al:train} 
\begin{algorithmic}[1]
 \Require\\ 
 $x\coloneqq $ linguistic features; $c\coloneqq $ speaker code; $y\coloneqq $ mel-spectrogram;\\
 Initial generator parameter $\theta_G$ and initial discriminator parameter $\theta_D$;\\
 A well-trained WaveNet model $W$ and $\theta_W$ is fixed;\\
 batch size $m$, learning rate $\eta$, the gradient penalty coefficient $\lambda$,  weight for adversarial loss $r_D$, weight for DML loss $r_W$, generator warming up iterations $n1$, basic adversarial training iterations $n2$, number of total iterations $n3$.
 
\Begin step 1: warming up generator 
 \setcounter{ALG@line}{0} 
 \For{epoch = $1, \cdots, n1$}
 \For{training data in $(x,c,y)$} 
 \State{generate $\hat{y}$ from the generator} 
 \begin{center}{ $\hat{y} = G(x,c)$}\end{center}
 \State{update $\theta_G$ using MSE criterion:}
 \begin{center} {$\theta_G \gets \theta_G - \eta_G\nabla_{\theta_G}L_{MSE}(y,\hat{y})$}\end{center}
\EndFor 
 \EndFor 
\Endd 
\Begin step2: adversarial training
 \setcounter{ALG@line}{0} 
 \For{epoch = $n1, \cdots, n2$}
 \For{training data in $(x,c,y)$} 
 \For{$i = 1, \cdots, m$}
 \begin{flushleft} {\qquad\qquad $\hat{y} = G(x,c)$}\end{flushleft}
 \begin{flushleft} {\qquad\qquad $\tilde{y} = \epsilon y + (1-\epsilon \hat{y}), \quad \epsilon \in U[0,1]$}\end{flushleft}
 \begin{flushleft} {\qquad\qquad $L_D^{(i)} = D(y)-D(\hat{y})+\lambda (\left \|\nabla_{\widetilde{y}} D(\widetilde{y}) \right \|_2-1)^2$}\end{flushleft}
 \EndFor 
 \State {update $\theta_D$ while fixing $\theta_G$: 
 \begin{center}{ $\theta_D \gets \theta_D - \eta_D\nabla_{\theta_D}\frac{1}{m}\sum_{i=1}^{m}L_D^{(i)}$}\end{center}}
 \State{update $\theta_G$ using both MSE and adversarial criterion:}
 \begin{center} $L_{ADV}=\frac{1}{m}\sum_{i=1}^{m}D(G(x,c))$ \end{center}
 \begin{center} $\theta_G \gets \theta_G - \eta_G\nabla_{\theta_G}(L_{MSE}(y,\hat{y}) +\gamma_{D}L_{ADV})$
 \end{center}
\EndFor 
 \EndFor 
\Endd 
\Begin {step 3: fine tuning the generator by utilizing WaveNet loss.} 
 \setcounter{ALG@line}{0} 
 \For{epoch = $n2, \cdots, n3$}
 \For{training data in $(x,c,y)$} 
 \State{generate $\hat{y}$ and update $\theta_D$ following step 2.}
 \State{upsampling $\hat{y}.$}
 \State{generate $\hat{s}$ from the well-trained WaveNet model:} 
 \begin{center}{ $\hat{s} = W(\hat{y}, c)$}\end{center}
 \State{update $\theta_G$ with DML loss from WaveNet:}
 \[\theta_G \gets \theta_G - \eta_G\nabla_{\theta_G}(L_{MSE}(y,\hat{y})
 +\gamma_{D}L_{ADV}+\gamma_{W}L_{DML}(s,\hat{s})) \]
\EndFor 
 \EndFor 
\Endd 
\end{algorithmic}
\end{algorithm}

\subsection{Time resolution adjustment}
\label{time}
During the training of the multi-speaker acoustic model, there are two instances where we need to pay attention to \textit{time resolution} problems. The first is when the mel-spectrograms are input to the WaveNet vocoder. The other is when DML loss is applied for generator optimization.

When acoustic features are transformed into speech samples, conventional parametric vocoders always use interpolation inside frames to recover the audio sampling points. Since different sampling points may share the same acoustic features, in existing studies related to WaveNet, several approaches have been proposed to align the input conditional features with the speech samples. 

When acoustic features are transformed into speech samples in the WaveNet vocoder, Oord et al.\ used a trainable transposed convolutional network to upsample the time resolution of the conditional acoustic features. Deep Voice 2 applied a stack of bidirectional quasi-recurrent neural networks and Tamamori et al.\ simply duplicated the conditional acoustic feature vector of each frame. In our work, we use trainable transposed convolutional layers to align mel-spectrograms and speech samples for the WaveNet vocoder as in~\cite{van2016wavenet}.

\begin{figure}[tb]
\begin{center} 
\includegraphics[clip, trim=3cm 2cm 2.5cm 2.5cm, width=0.5\textwidth]{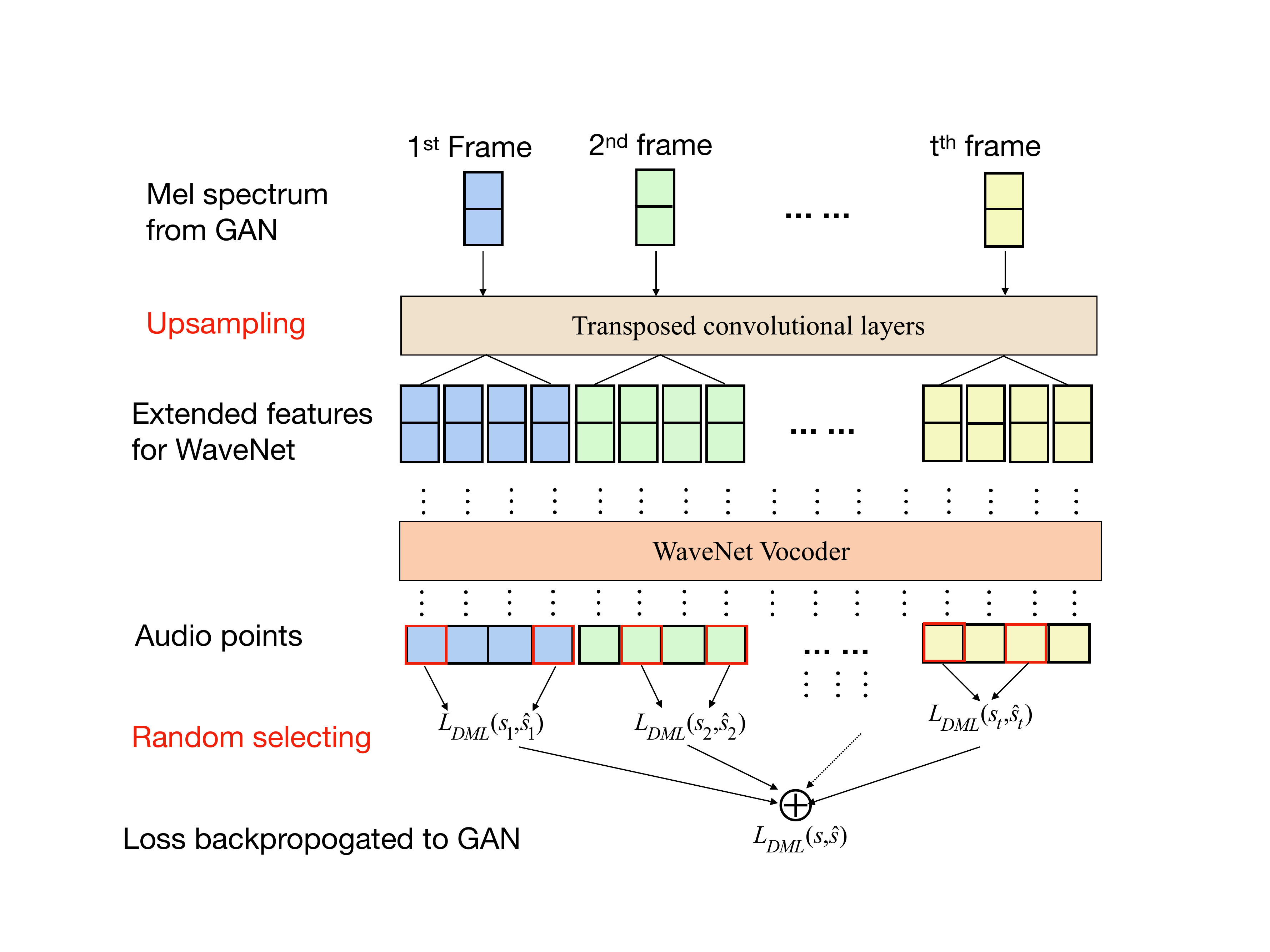}
\end{center}
\caption{Time resolution adjustment of conditional acoustic features. One frame includes four waveform audio points. Transposed convolutional layers are used to upsample the conditional acoustic features. The DML loss was computed using randomly selected waveform audio points within each frame.}
\label{img:time_res}
\end{figure}

When the well-trained WaveNet vocoder is used for the proposed generator optimization, it would be time-consuming to calculate the DML loss along all the waveform audio samples within the same frame. As shown in Fig.~\ref{img:time_res}, in order to improve computational efficiency, we randomly select a part of the waveform audio points within each frame and back-propagate their averaged DML loss to the generator for acoustic model optimization.

\section{Experimental Setup}
\label{sec:exp}
We used six speakers (awb, bdl, clb, ksp, rms, and slt) from the CMU-ARCTIC database for multi-speaker training. Two speakers (clb and slt) are female and the others are male. For each speaker, 1000 utterances were used for training. Their speech waveforms have a sampling frequency of 16 kHz and a 16-bit PCM format. The six speakers read out the same set of utterances. Linguistic labels were generated by Festival TTS and consist of 376-dimensional binary vectors and 5-dimensional duration information. The linguistic features are normalized by the min-max rule. Speaker codes consist of seven dimensions, where six dimensions represent speaker identity difference in one-hot format and the other dimension denotes gender. The speaker codes are input to the first layer of both the generator and discriminator as auxiliary features. For the WaveNet vocoder, the speaker codes are first input to a fixed-size embedding layer and then converted to an input format compatible with WaveNet. None of the utterances in the testing set appear in either the training or development sets.

As acoustic features, 80-dimensional static mel-spectrograms are adopted in our experiment. To compute mel-spectrograms, we first perform a short-time Fourier transform (STFT) on audios using a 15-ms frame size, 5-ms frame shift, and a Hann window function. Then we transform the STFT magnitude spectrum to the mel scale using an 80-channel mel-filterbank that ranges from 125 Hz to 7.6 kHz, followed by log dynamic range compression. Prior to the log compression, the filterbank output magnitudes are clipped to a minimum value of 0.01 in order to limit the dynamic range in the logarithmic domain. The mel-spectrograms are then normalized to have zero-mean unit variance.
 
We used six bidirectional SRU layers for acoustic modeling and three feed-forward layers for the discriminator. In the generator, each layer has 512 hidden nodes, and in the discriminator, each layer has 128 hidden nodes. The ReLU activation function is utilized in the SRU cell. A stochastic gradient descent (SGD) optimizer was used as the optimizer for both the generator and discriminator. Learning rate was initialized to 0.01 for the generator and 0.001 for discriminator along with exponential decays corresponding to the number of training epochs. 

To implement the WaveNet model, we referenced ~\cite{paine2016fast} and adopted a modified version of the WaveNet architecture. Instead of predicting discretized buckets with a softmax layer, we followed Tacotron 2 and Parallel WaveNet and used a 10-component mixture of logistic distributions to generate 16-bit samples at 16 kHz. To compute the logistic mixture distribution, the WaveNet stack output was passed through a ReLU activation, followed by a linear projection to predict parameters (mean, log scale, mixture weight) for each mixture component. We adopted 24 dilated convolution layers grouped into four dilation cycles. The dilation rate of the $k$-th layer was set to $2^{k\pmod{6}}$, where $k\in[0,1, 2 \cdots23]$. Finally, 24 residual blocks were connected. The number of channels of (dilated) causal convolution and $1 \times 1$ convolution in the residual block were set to 512. The number of $1\times 1$ convolution channel between skip-connection and output layer was set to 256. We used three transposed convolutional layers for up-sampling. The Adam algorithm ~\cite{kingma2014adam} was used for the optimization, and its learning rate was initialized to 0.001 and scheduled carefully with a scheme similar to~\cite{vaswani2017attention}. Other parameters in the Adam optimizer were set as $\beta_1 = 0.9, \beta_2=0.999, \epsilon=1.0e^{-8}$. We also maintained an exponentially weighted moving average of the network parameters over update steps with a decay of $0.9999$. A GeForce GTX 1080 was used for training. It took about a week to train a high-quality multi-speaker WaveNet vocoder and eight minutes to synthesize ten seconds of speech. 
When updating the generator using the DML loss back-propagated from the trained WaveNet Vocoder, we randomly chose half of the sampling points in each frame to efficiently calculate the DML loss. 
$\gamma_D$ was set equal to $E(L_{MGE})/E(L_{ADV})$, and $E(\cdot)$ represented expectation value. $\gamma_W$ was fixed as 0.0001. 

\section{Experimental Evaluation}
\label{sec:res}
We compared the performance of the following configurations based on a listening test:
\begin{enumerate}
\item Baseline: Acoustic model trained using $L_{MSE} (y,\hat{y})$ as a criterion.
\item GAN: Acoustic model trained using $L_{MSE} (y,\hat{y}) + \gamma_D L_{ADV} (\hat{y})$ as a criterion.
\item GAN\textsuperscript{W}: Acoustic model trained using $L_{MSE} (y,\hat{y}) + \gamma_D L_{ADV} (\hat{y}) + \gamma_W L_{DML} (y,\hat{y})$ as a criterion.
\item WGAN-GP: Acoustic modeling trained using $L_{MSE} (y,\hat{y})+\gamma_D{L_{ADV}}(\hat{y})$ as a criterion. WGAN-GP was also used. 
\item WGAN-GP\textsuperscript{W}: Acoustic model trained using $L_{MSE} (y,\hat{y}) + \gamma_D L_{ADV} (\hat{y}) + \gamma_W L_{DML} (y,\hat{y})$ as a criterion. WGAN-GP was also used. 
\item Analysis by synthesis (AbS): Synthetic speech generated by a WaveNet vocoder using ground-truth mel-spectrograms.
\item Natural: Natural speech.
\end{enumerate}
Note that systems 1 to 5 are TTS systems and use SRU as basic architectures for acoustic models, as described earlier. Also note that all the above TTS systems and analysis by synthesis use the same WaveNet vocoder. The differences are how the local condition parameters of the WaveNet vocoder, that is, mel-spectrogram, are predicted. 

\subsection{Evaluation methodology}
\label{subsec:eva_meth}
For the listening test, we selected 20 utterances from the testing set of each speaker and generated sets of synthetic speech corresponding to the above experimental systems. Each experimental system had 20 utterances, so 20 utterances $\times$ 6 speakers $\times$ 7 = 840 samples that needed to be evaluated in total. Crowdsourced perceptual evaluation was carried out to evaluate naturalness as well as speaker similarity of generated speech. In the crowdsourcing test, we evaluated each sample ten times to alleviate personal bias. The testing samples were divided into different evaluation sets. Each set consisted of three utterances generated by seven different systems. Therefore, there were 42 utterances to be evaluated in each set: 21 for naturalness and 21 for similarity. We then collected 400 sets to cover all 840 samples (400 = 840$\times$10/21). This guarantees at least 40 unique listeners, since we limited the maximum number of sets per crowdsourced participant to ten. The actual number of listeners who participated in our test was 42.

To evaluate naturalness, listeners were asked to ignore the meaning of the sentence and concentrate only on rating how natural the speech sounded on a five-point scale: 
\begin{enumerate}
\item completely unnatural
\item mostly unnatural
\item equally natural and unnatural
\item mostly natural
\item completely natural
\end{enumerate}

For speaker similarity, listeners were asked to ignore the meaning of the sentence and concentrate only on rating the speaker identity. Synthetic speech samples and the corresponding natural sound were presented in pairs at every turn and listeners were asked to judge whether the two samples were from the same or different speaker(s). The scale for speaker similarity was judged on a four-point scale: 
\begin{enumerate}
\item same speaker, absolutely sure
\item same speaker, not sure 
\item different speaker, not sure
\item different speaker, absolutely sure
\end{enumerate}

\subsection{Evaluation results and analysis}
\label{subsec:res_ana}
\begin{figure}[tb]
\begin{center} 
\includegraphics[width=8cm]{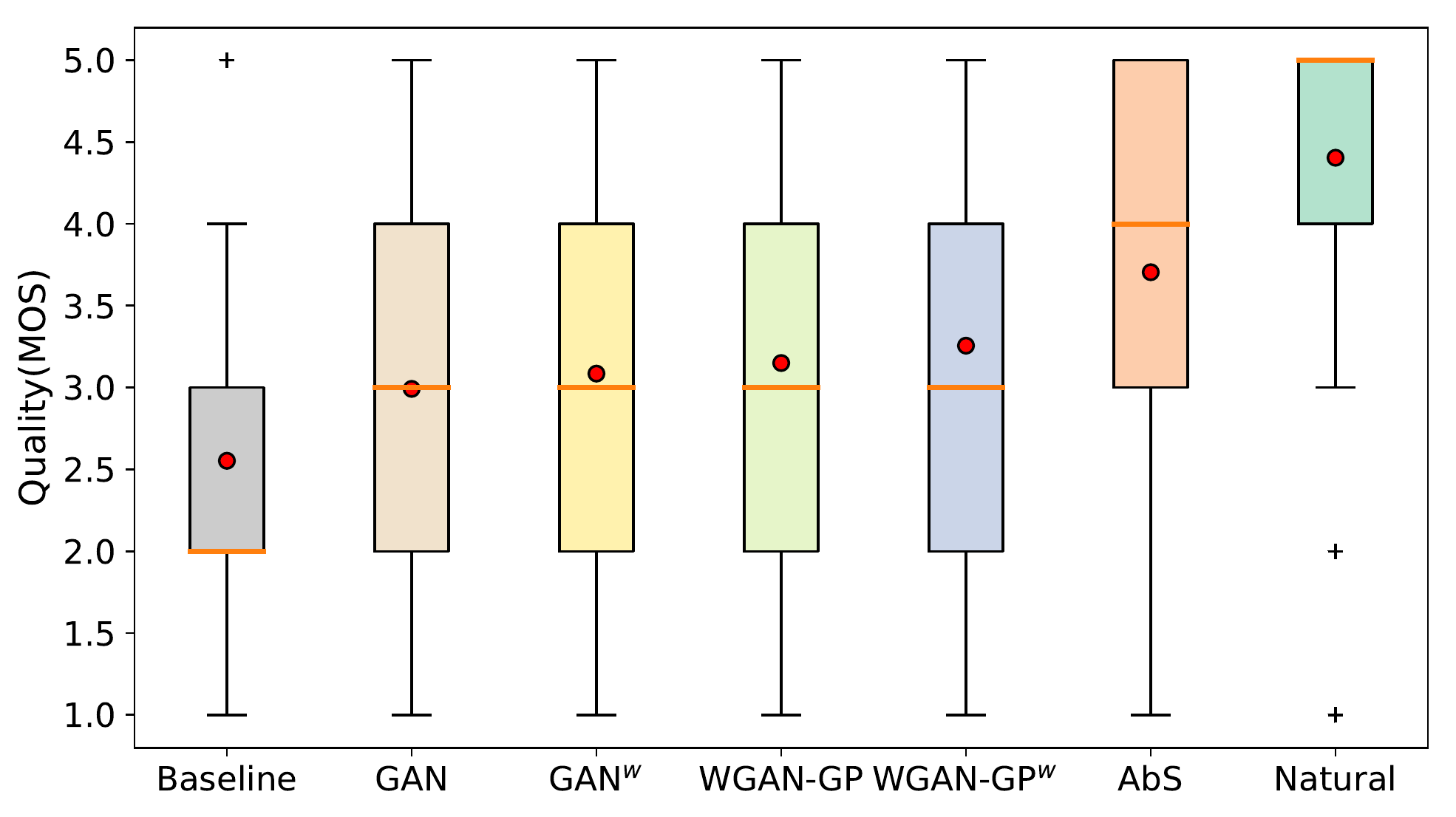}
\end{center}
\caption{Box plots on naturalness evaluation results. Red dots represent the mean of each group averaged across all speakers.}
\label{img:qua}
\end{figure}

\begin{table*}[t]%
\setbox0\hbox{\verb/\documentclass/}%
\caption{Statistical significance analysis using $t$-tests with Holm-Bonferroni correction in terms of quality judgment.}
\label{tb:t-qua}
\begin{center}
\small
\begin{tabular}{llllllll}
\hline
System& Baseline&GAN&GAN\textsuperscript{W}&WGAN-GP&WGAN-GP\textsuperscript{W} & AbS \\
\hline 
GAN&<2e-16&-&-&-&-&- \\

GAN\textsuperscript{W}& < 2e-16&0.05206&-&-&-&- \\

WGAN-GP& < 2e-16&0.00028&0.19850&-&-&- \\

WGAN-GP\textsuperscript{W}& < 2e-16&1.2e-06&0.01916&0.24092&-&- \\

AbS & < 2e-16&< 2e-16&< 2e-16&<2e-16&<2e-16 &- \\

Natural& < 2e-16&< 2e-16&< 2e-16&< 2e-16&< 2e-16& < 2e-16 \\

\hline
\end{tabular}%
\end{center}
\end{table*}

Fig.~\ref{img:qua} shows the box plots for the naturalness evaluation results averaged across all speakers. Table~\ref{tb:t-qua} shows statistical significance. From these, we can see that four GAN-based experimental groups (GAN, GAN\textsuperscript{W}, WGAN-GP, WGAN-GP\textsuperscript{W}) outperform the baseline significantly. Upper quartiles and mean opinion scores of the four GAN-based groups are much higher than those of the baseline, although their lower quartiles are quite similar to the baseline. Note that all the systems (apart from natural speech) use the same WaveNet vocoder. Hence, this also indicates that the quality of WaveNet synthetic speech is affected by the local condition parameters and that the ones predicted by the GAN-based acoustic models sound more natural than those by the baseline. We also see that WGAN-GP systems (WGAN-GP, WGAN-GP\textsuperscript{W}) are better than the original GAN system. The use of DML loss alone did not bring statistically significant improvements, but it obviously reduced $p$-values (see Table~\ref{tb:t-qua}), and hence a combination of WGAN-GP and the DML loss resulted in the highest scores among the TTS systems and was significantly better than GAN and GAN\textsuperscript{W} ($p<0.05$).

Compared with the natural speech and AbS, all TTS methods have obvious gaps. There is also a gap between the AbS samples and natural speech. This indicates that our multi-speaker TTS systems do not sound as good as natural speech yet, and the multi-speaker WaveNet vocoder itself does not sound as good as natural speech either, even if it uses the ground-truth mel-spectrogram. In other words, both the neural vocoder and the acoustic model have room for further improvement. 

\begin{figure}[tb]
\centering
\includegraphics[width=8cm]{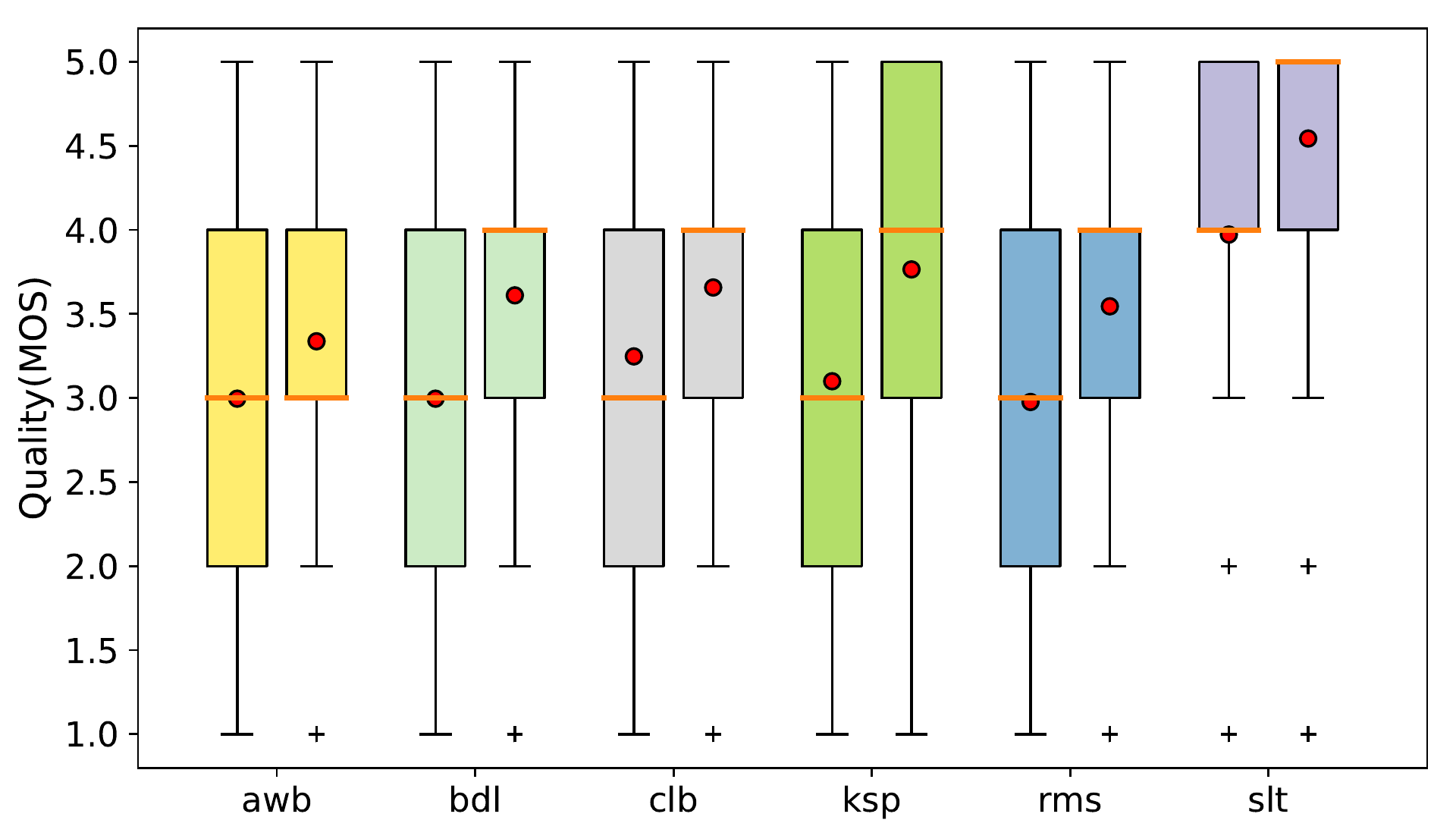} \\
\caption{Box plots of the MOS scores of six speakers. Left: WGAN-GP\textsuperscript{W} system. Right: AbS system.}
\label{img:mos_sp}
\end{figure}

Through our experiments, we found that the quality of our synthetic speech varied speaker by speaker. Fig.\ref{img:mos_sp} shows box plots of the MOS scores of the best WGAN-GP\textsuperscript{W} system and the AbS system of the six speakers. The left box plot shows the results of the WGAN-GP\textsuperscript{W} system and the right box plot shows those of the AbS system for each speaker. Interestingly, the quality of synthetic speech varied speaker by speaker, and there is a very large gap between speaker SLT and the other speakers. This implies that we need a more generalized model that can handle multiple speakers better and can reproduce the differences between speakers more precisely. 

\begin{figure}[tb]
\begin{center} 
\includegraphics[width=8.2cm]{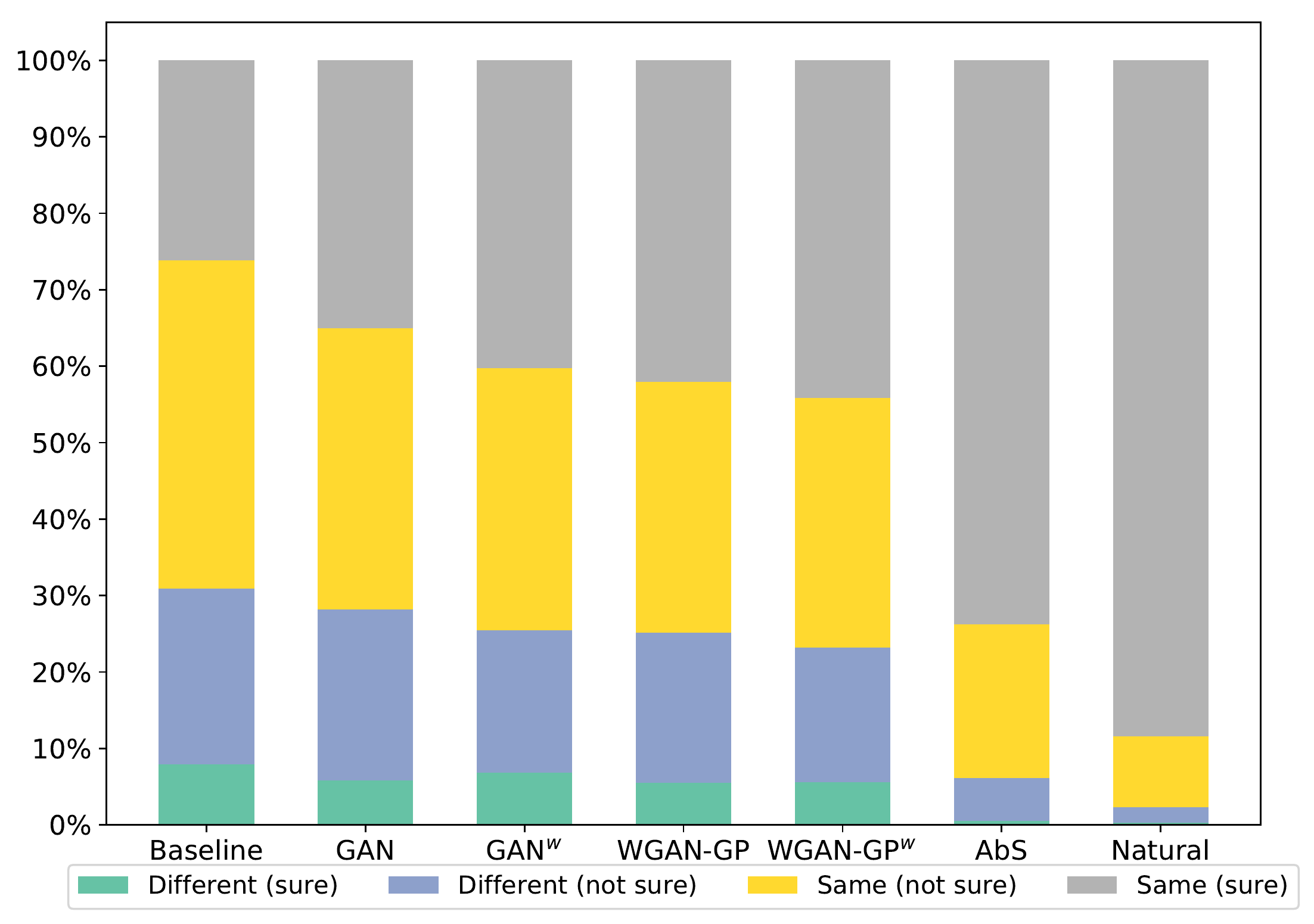}
\end{center}
\caption{Similarity results averaged across all speakers.}
\label{img:sim}
\end{figure}

\begin{table*}[t]%
\setbox0\hbox{\verb/\documentclass/}%
\caption{Statistical significance analysis using $t$-tests with Holm-Bonferroni correction in terms of speaker similarity judgment.}
\label{tb:t-sim}
\begin{center}
\small
\begin{tabular}{llllllll}
\hline
Systems& Baseline&GAN&GAN\textsuperscript{W}&WGAN-GP&WGAN-GP\textsuperscript{W} & AbS \\
\hline 
GAN&0.43810&-&-&-&-&- \\

GAN\textsuperscript{W}& 0.28401&1.00000&-&-&-&- \\

WGAN-GP& 0.00426&0.31593&0.47565&-&-&- \\

WGAN-GP\textsuperscript{W}& 0.00044&0.11438&0.28401&1.00000&- &- \\

AbS&<2e-16&<2e-16&<2e-16&<2e-16&<2e-16&- \\

Natural& <2e-16&<2e-16&<2e-16&<2e-16&<2e-16&<4.2e-06 \\

\hline
\end{tabular}%
\end{center}
\end{table*}

The similarity evaluation results are shown in Fig.\ref{img:sim} and the t-test results for similarity are shown in Table~\ref{tb:t-sim}. The trend of the similarity tests is very similar to that of the naturalness. The WGAN-based systems outperform the baseline, and we can clearly see that the portions of "Same" (yellow and gray) have been increased. The proposed systems using a combination of WGAN-GP and DML loss achieved more apparent preference in terms of "Same, absolutely sure". Likewise in the quality evaluation, we can see a gap between TTS systems and WaveNet analysis-by-synthesis systems as well as between WaveNet analysis-by-synthesis systems and natural speech. 

\begin{figure}[tb]
\begin{center} 
\includegraphics[width=8cm]{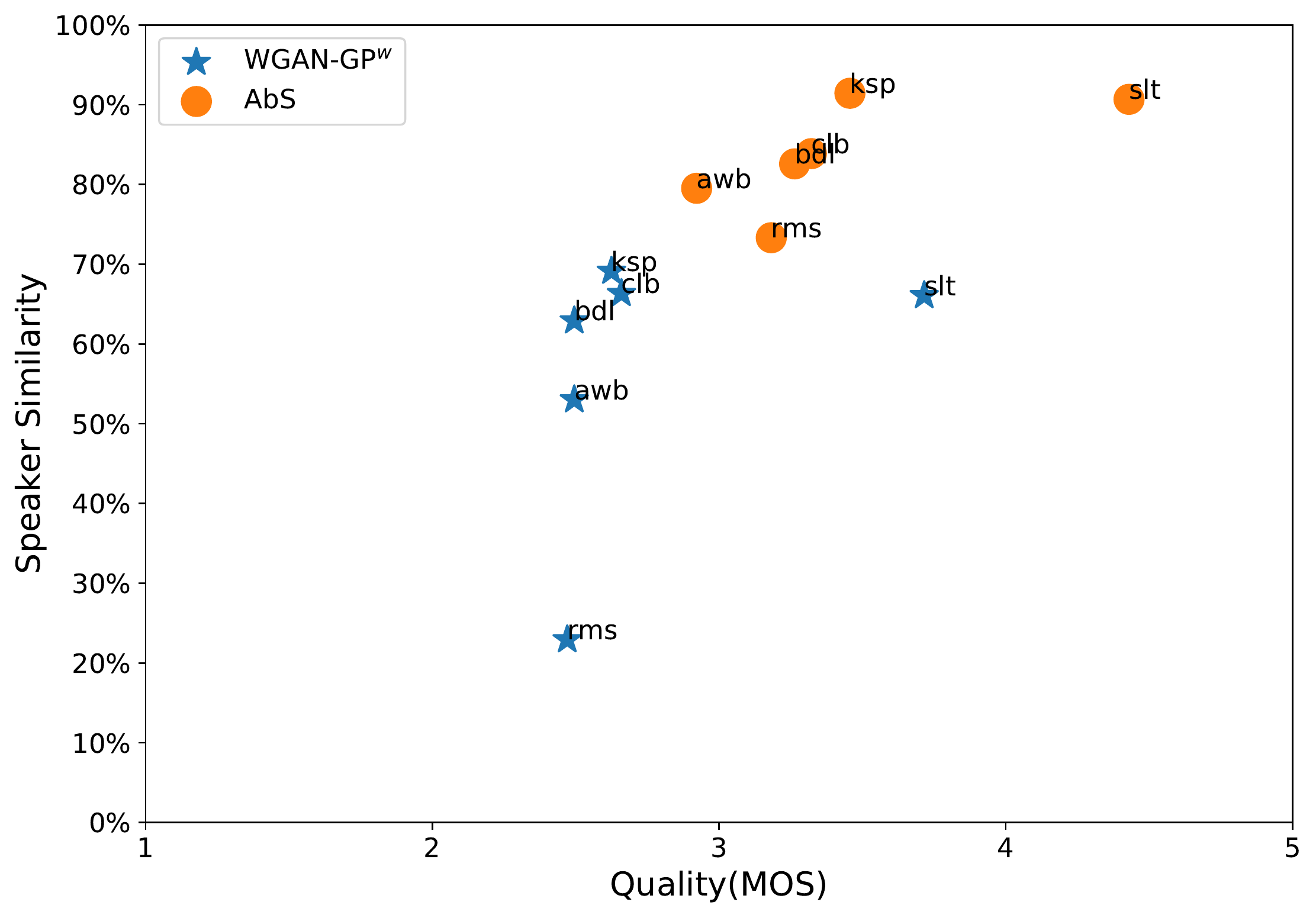}
\end{center}
\caption{Scatter plot matching naturalness and similarity scores for each speaker in system WGAN-GP\textsuperscript{W} and AbS. The similarity score is defined as the added percentage of `same (not sure)' and `same (sure)' scores.} 
\label{img:scatter}
\end{figure}

Fig.~\ref{img:scatter} shows a scatter plot matching naturalness and similarity scores of the best WGAN-GP\textsuperscript{W} system and AbS system of six speakers. Interestingly, the speaker similarity scores also significantly varied speaker by speaker, and speaker RMS had a very low speaker similarity score. Our next step is to investigate why a few speakers had lower speaker similarity.

\section{Conclusion}
\label{sec:con}
This paper investigated how we should train the acoustic model that predicts the local condition parameters to be used by neural vocoders. Specifically, we looked into conditional GANs or WGAN-GP to reduce the mismatched characteristics between natural and generated acoustic features. We also extended the GAN frameworks and used the discretized mixture logistic loss of a well-trained WaveNet along with mean squared error and adversarial losses as parts of the objective functions. These new objective functions were evaluated in multi-speaker speech synthesis that uses the WaveNet vocoder. Experimental results show that acoustic models trained with the WGAN-GP framework using back-propagated DML loss achieved the highest subjective evaluation scores in terms of both quality and speaker similarity.

Our future work will investigate why some speakers have lower quality of synthetic speech or lower similarity. We will also perform larger scale experiments using more speakers.

\bibliographystyle{IEEEtran}
\bibliography{main}

\EOD

\end{document}